%% file: main-lncs.tex
\documentclass[runningheads]{llncs}
\usepackage[T1]{fontenc}
\usepackage{amsmath}
\usepackage{amssymb}
\usepackage{graphicx}
\usepackage{booktabs}
\usepackage{array}
\usepackage{xcolor}
\usepackage{colortbl}
\newcommand{\rowrule}{\arrayrulecolor{black!25}\specialrule{0.3pt}{1.5pt}{1.5pt}\arrayrulecolor{black}}
\usepackage{tikz}
\usetikzlibrary{positioning,calc}
\usepackage{placeins}
\usepackage{float}
\usepackage[colorlinks=true,linkcolor=blue,citecolor=blue,urlcolor=blue]{hyperref}

\newcommand{\resmethw}{5.2cm}\newcommand{\resdatw}{1.3cm}\newcommand{\resheadw}{2.81cm}
\tikzset{dagfig/.style={x=0.9cm}, dagb/.style={xshift=6.6cm, x=0.75cm}}

\begin{document}
\title{GFlowQEC: Generative Flow Networks that Learn to Sample Quantum LDPC Codes\thanks{This work is funded by national funds through FCT -- Foundation for Science and Technology, I.P., within the scope of the research unit UID/00326 -- Centre for Informatics and Systems of the University of Coimbra, \url{https://doi.org/10.54499/UID/00326/2025}. This work is funded by national funds through FCT -- Fundação para a Ciência e a Tecnologia, I.P., and, when eligible, co-funded by EU funds under project/support UID/50008/2025 -- Instituto de Telecomunicações, with DOI identifier \url{https://doi.org/10.54499/UID/50008/2025}.}}
\titlerunning{GFlowQEC}
\author{Nuno Batista\inst{1} \and Gabriel Falcão\inst{2,3} \and Penousal Machado\inst{4}}
\authorrunning{N. Batista et al.}
\institute{Department of Informatics Engineering, University of Coimbra, Coimbra, Portugal \and
Instituto de Telecomunicações, Department of Electrical and Computer Engineering, University of Coimbra, Coimbra, Portugal \and
New York University Abu Dhabi, Abu Dhabi, United Arab Emirates \and
University of Coimbra, CISUC/LASI -- Centre for Informatics and Systems of the University of Coimbra, Department of Informatics Engineering\\[1ex]
\email{nuno.batista@uc.pt}, \email{gff9537@nyu.edu}, \email{machado@dei.uc.pt}}
\maketitle
\begin{abstract}
Choosing a quantum error-correcting code for a real device means scoring candidates with circuit-level simulations so slow that only a few hundred can be run.
So every evaluation counts, and a search is judged by how many good codes it finds per paid evaluation.
Quantum LDPC codes built from a few group elements protect many logical qubits, but good ones are rare in a huge space.
We introduce GFlowQEC, to our knowledge the first generative flow network (GFlowNet) for quantum code design.
It builds a code one group element at a time and learns to propose codes in proportion to a power of their score.
We train it once on scores from earlier searches, freeze it, and let it propose lifted-product codes with $180$ physical qubits, scored by their distances.
Ranking ten of its draws per paid call, it finds more than twice as many new top-scoring codes (inequivalent to any it was trained on) as tuned parallel tempering, even when tempering starts from GFlowQEC's training codes.
It finds more than three and a half times as many as a supervised screen ranking ten random candidates per call.
When both rank a hundred, the screen beats GFlowQEC, and does best ranking GFlowQEC's draws.
Soft Q-learning, which shares the GFlowNet's optimal policy, does as well at one noise bias and better at the other, so the gain does not depend on the GFlowNet's training loss.
Trained proposers can make a budget of expensive evaluations go further when good codes are rare.
\end{abstract}
\input{sections/01-introduction}
\input{sections/02-related-work}
\input{sections/03-background}

\input{sections/04-benchmark}
\input{sections/05-policy}
\input{sections/08-evaluation-protocol}
\input{sections/09-results}
\input{sections/10-discussion}
\input{sections/11-future-work}
\begin{credits}
\subsubsection{\ackname}
The authors proposed most of the ideas, designed and directed the experiments and carried out the literature review. The large language models Claude Fable 5.1, Claude Opus 5.5 and Claude Sonnet 5.5 (Anthropic) wrote almost all of the code and helped summarise the literature, analyse the results, and write and edit the text, always under the authors' guidance. The authors reviewed the text and take full responsibility for its content.
\subsubsection{\discintname}
The authors have no competing interests to declare that are relevant to the content of this article.
\end{credits}
\bibliographystyle{splncs04}
\def\doi#1{\href{https://doi.org/#1}{doi:\nolinkurl{#1}}}
\bibliography{refs}
\clearpage
\end{document}

%% file: sections/01-introduction.tex
\section{Introduction}
\label{sec:intro}
Quantum low-density parity-check (LDPC) codes can protect many logical qubits in a single block of physical qubits~\cite{mackay2004,breuckmann2021}. Many of them are specified by a short list of elements of a finite group, and since every list gives a valid code~\cite{lin2023,panteleev2022lp}, designing one amounts to a search over lists. That search is hard for two reasons. The space is enormous: the $[[300,60,14]]$ mitten code of Bhardwaj et al.~\cite{bhardwaj2026} is a list of ten elements of a group with sixty elements, and lists of that length give up to about $3 \times 10^{8}$ inequivalent codes. And every candidate has to be scored, either by computing its distance or by simulating how often it fails, which takes from about a second to a few minutes per code. A search therefore only ever sees a sliver of the space.

Our motivating setting is choosing a code for a particular device.
There, each candidate is scored by a circuit-level simulation, with faulty measurements and a real decoder, which takes minutes to hours per code, so a search scores only hundreds~\cite{yan2026,bravyi2024,bhardwaj2026}.
Since the best code can also change with the device's noise and connectivity~\cite{bravyi2024,berthusen2025}, a cheaper score can only shortlist candidates for the expensive simulation, so a search should deliver many good candidates for its budget rather than a single best one~\cite{bengio2021}.
Because methods cannot yet be compared fairly at that price, we compare them where scoring is cheap enough to run each method many times, but we keep the unit of cost that matters on a device: the paid call, that is, one call to the scoring routine (the oracle) for a code the run has not scored before.

Bayesian optimisation, reinforcement learning, filtered uniform sampling and large language models (LLMs) have already been used to search code families of this kind, and Table~\ref{tab:tasks} compares these searches with ours.
None of them, however, trains a model to sample codes in proportion to their quality.

Learning can turn a fixed budget into more good codes in two ways.
A screen draws random candidates and pays only for the ones a learned model ranks highest, whereas a proposer learns to draw good candidates directly.
When good codes are rare, the difference matters: a screen has to look at many random candidates for every one it keeps, while a proposer builds that rejection into its own draws.
Generative flow networks (GFlowNets)~\cite{bengio2021,bengio2023} are proposers of this kind. A GFlowNet builds an object step by step and is trained so that, at its optimum, it produces each object with probability proportional to a reward, which means one model yields many distinct good objects where reward maximisation would converge on one~\cite{bengio2021}. To our knowledge, no GFlowNet has yet been used to design quantum error-correcting codes.

This paper makes two contributions.
\begin{itemize}
\setlength{\itemsep}{0pt}
\item We introduce GFlowQEC, a GFlowNet whose policy fills in the list one group element per step and is trained with trajectory balance~\cite{malkin2022} to produce each list with probability proportional to a power of its score.
\item We test this policy, as a proof of concept, on lifted-product codes with $180$ physical qubits, where it is trained once on scores stored from earlier searches, frozen, and compared with tuned parallel tempering, a supervised screen and soft Q-learning at matched budgets of paid calls.
\end{itemize}
\begin{table*}
\caption{Prior code searches compared with this paper. LP stands for lifted product and BB for bivariate bicycle codes, two-block codes built from two sums of elements of $Z_l \times Z_m$, the pairs of integers mod $l$ and mod $m$ (Bravyi et al.~\cite{bravyi2024} allow only pairs with one entry zero).}
\label{tab:tasks}
\footnotesize
\setlength{\tabcolsep}{3pt}
\begin{tabular}{@{}>{\raggedright\arraybackslash}p{3.0cm}>{\raggedright\arraybackslash}p{1.9cm}>{\raggedright\arraybackslash}p{2.0cm}>{\raggedright\arraybackslash}p{2.0cm}>{\raggedright\arraybackslash}p{1.2cm}>{\raggedright\arraybackslash}p{1.0cm}@{}}
\toprule
Work & Family & Score & Returns & Matched budget & Exact distances \\
\midrule
Chengyu et al.~\cite{chengyu2026}, Bayesian optimisation & BB, any weight & Rate $k/n$ and error rate & one code per weighting & yes & no \\
\rowrule
Cruz-Benito et al.~\cite{cruzbenito2026}, LLM & BB & $kd^2/n$ & a catalogue & partly & some \\
\rowrule
Liu and Marquardt~\cite{liu2026}, LLM & LP & Rate $k/n$ and error rate & an archive & no & no \\
\rowrule
Bhardwaj et al.~\cite{bhardwaj2026}, filtered sampling & LP, non-abelian & Distance upper bounds & one code per group & no & some \\
\rowrule
Sella et al.~\cite{sella2026}, reinforcement learning & BB as defined by Bravyi et al.~\cite{bravyi2024} & Decoder success rate & best code so far & no & no \\
\rowrule
Freire et al.~\cite{freire2025}, three searches & Hypergraph product & Erasure failure rate & one best code & yes & not stated \\
\midrule
This paper, GFlowNet policy & LP, non-abelian & Failure exponent, Eq.~\eqref{eq:exponent} & codes proposed and ranked by one frozen policy & yes & some \\
\bottomrule
\end{tabular}
\end{table*}

At $180$ physical qubits the frozen policy draws ten candidates for each paid call and pays for the ones it ranks most probable.
Counting only codes inequivalent to those it was trained on, it found more than twice as many top-scoring codes per paid call as tuned parallel tempering (tuned PT), $2.4$ and $2.8$ times as many at the two noise biases we test, where bit flips are ten times likelier than phase flips or the reverse.
Started from the same stored codes, tuned PT did not close the gap, so the lead is not just access to the stored data.
It also found more than three and a half times as many as a supervised screen built on the same network and trained on the same scores.
Because only about one random code in a hundred is top-scoring here, a screen shown ten random codes per paid call rarely sees one.
Shown a hundred random codes per paid call, the screen found more than the policy, and the largest count of all came from a screen that ranked the policy's own draws.
Soft Q-learning trained in the same way, which shares the GFlowNet's optimal policy on our construction~\cite{tiapkin2024}, did as well at one bias and better at the other, so the gain does not depend on the GFlowNet's training loss.

A longer version with all settings and further results is available with the code and data at \url{https://doi.org/10.5281/zenodo.23922222}.

%% file: sections/02-related-work.tex
\section{Related Work}
\label{sec:related}
\paragraph{Code search.}
Simple baselines are strong in this setting: Freire et al.~\cite{freire2025}, who compare methods at a matched budget under one oracle as we do, find on hypergraph-product codes that a random walk matches simulated annealing and reinforcement learning.
Beyond Table~\ref{tab:tasks}, LLM agents search under circuit-level noise~\cite{yan2026} and over balanced-product codes~\cite{qian2026}. Reinforcement learning has glued small codes into codes of up to $20$ physical qubits with one logical qubit~\cite{su2025} and has found a code on nine physical qubits together with its encoder, with the noise bias as an input~\cite{olle2024}.
We do not reimplement the LLM searches~\cite{cruzbenito2026,liu2026,yan2026,qian2026}, because their oracles and budgets differ from ours and their models may have seen published codes of these families during pretraining.

\paragraph{GFlowNets.}
On a tree, where every object has a single construction path, GFlowNet training is equivalent to entropy-regularised reinforcement learning such as soft Q-learning~\cite{tiapkin2024}.
Conditional GFlowNets take a parameter of the reward as an input, such as the reward's temperature in Logit-GFN~\cite{kim2024b} or the noise bias in GFlowQEC.
Training a GFlowNet only on stored rewards, with no reward model and no new oracle call, is an established setting~\cite{chen2026tdgfn}. Like the plain baseline of that work, GFlowQEC trains directly on its stored codes, and it adds conditioning on the bias and a deployment that ranks its own draws.
GFlowNets have also been applied to quantum circuits: FlowQ-Net~\cite{dai2025} builds them gate by gate and lists quantum error-correcting code synthesis among its future applications, and QFlowNet~\cite{koo2026} applies a GFlowNet to unitary synthesis.

%% file: sections/03-background.tex
\section{Background}
\label{sec:setting}

\subsection{Quantum LDPC Codes}
\label{sec:css}
An error on a qubit is a bit flip ($X$), a phase flip ($Z$) or both ($Y$). A Calderbank-Shor-Steane (CSS) code~\cite{calderbank1996,steane1996} is specified by two binary matrices $H_X$ and $H_Z$, each with one column per physical qubit, $n$ in all. Each row is a check, that is, a parity measurement on the qubits where the row has a 1, and the pair of matrices defines a valid code exactly when
\begin{equation}
  H_X H_Z^{\mathsf T} = 0 \pmod 2 .
  \label{eq:css}
\end{equation}

The number of logical qubits is
\begin{equation}
  k = n - \operatorname{rank} H_X - \operatorname{rank} H_Z ,
  \label{eq:k}
\end{equation}
with the ranks taken over $\mathbb{F}_2$, the field with two elements (arithmetic mod 2). A pattern of phase flips $e_Z \in \mathbb{F}_2^n$ marks the qubits that are hit, and its weight $|e_Z|$ is the number of such qubits. The $X$ checks detect the pattern when $H_X e_Z \neq 0$, and the $Z$ checks detect bit flips $e_X$ in the same way. Throughout, the subscript $X$ or $Z$ of $e$, $d$ and $p$ names the kind of flip, so $e_Z$, $d_Z$ and $p_Z$ all refer to phase flips.

Sums of rows of $H_Z$, read as phase flips, are the $Z$ stabilizers: they leave the code's states unchanged and are not seen by the $X$ checks (Eq.~\eqref{eq:css}). Every other undetected pattern changes the encoded information and is a logical operator, and the lightest logical operator of each kind defines the distance,
\begin{equation}
\begin{aligned}
  d_Z &= \min\{\, |e_Z| : H_X e_Z = 0,\; e_Z \notin \operatorname{rowspace} H_Z \,\}, \\
  d_X &= \min\{\, |e_X| : H_Z e_X = 0,\; e_X \notin \operatorname{rowspace} H_X \,\}, \\
  d &= \min(d_X, d_Z),
\end{aligned}
\label{eq:distance}
\end{equation}
where $\operatorname{rowspace} H$ is the set of sums of rows of $H$. A code with these parameters is written $[[n,k,d]]$. Its Tanner graph has one vertex for each qubit and each check, and an edge wherever a check's row has a 1. For a quantum LDPC code this graph is sparse~\cite{mackay2004}.

\subsection{Lifted-Product Codes over a Group}
\label{sec:families}
A code in our family is specified by a tuple $x$ of ten elements of a group $G$, and every search method in this paper proposes such tuples. We use $G = Z_6 \times S_3$, whose elements pair an element of $Z_6$, the integers mod 6 under addition, with one of the six orderings of three objects (the group $S_3$), combined entry by entry, so $G$ has $|G| = 36$ elements. A group is abelian when the order of a product does not matter, and $G$ is not. A sum such as $a = g_1 + g_2 + g_3$ is a set of group elements. The $|G| \times |G|$ binary matrices $\lambda(a)$ and $\rho(a)$, with rows and columns indexed by the elements of $G$, multiply by $a$ on the left and on the right: column $v$ of $\lambda(a)$ marks the elements $g v$, and column $v$ of $\rho(a)$ the elements $v g$, for every $g$ in $a$. The two differ only when $G$ is non-abelian. Finally, the sum $a^{*}$ replaces every element of $a$ by its inverse, so that $\lambda(a^{*})$ and $\rho(a^{*})$ are the transposes of $\lambda(a)$ and $\rho(a)$.

Our family extends the mitten codes of Bhardwaj et al.~\cite{bhardwaj2026} (their Definition~4), which are defined over non-abelian groups because over abelian groups their construction has distance at most $6$. Ten group elements and the identity $1$ form four sums of three distinct elements,
\begin{equation}
\begin{aligned}
  a_1 &= g_1 + g_2 + g_3, & a_2 &= 1 + g_4 + g_5, \\
  b_1 &= h_1 + h_2 + h_3, & b_2 &= 1 + h_4 + h_5.
\end{aligned}
\label{eq:entries}
\end{equation}

A base matrix has sums as its entries and becomes binary when each entry is replaced by its $|G| \times |G|$ matrix. The mitten codes are the lifted product~\cite{panteleev2022lp,panteleev2022good} of the row $(a_1 \; a_2)$ and the column $(b_1^{*} \; b_2^{*})^{\mathsf T}$, and in our convention the row enters through $\lambda$ and the column through $\rho$:
\begin{equation}
\begin{aligned}
  H_X &= \begin{pmatrix} \lambda(a_1) & 0 & \lambda(a_2) & 0 & \rho(b_1^{*}) \\ 0 & \lambda(a_1) & 0 & \lambda(a_2) & \rho(b_2^{*}) \end{pmatrix}, \\
  H_Z &= \begin{pmatrix} \rho(b_1) & \rho(b_2) & 0 & 0 & \lambda(a_1^{*}) \\ 0 & 0 & \rho(b_1) & \rho(b_2) & \lambda(a_2^{*}) \end{pmatrix},
\end{aligned}
\label{eq:lp}
\end{equation}
where $0$ is the $|G| \times |G|$ zero matrix. The code has $n = 5|G| = 180$ physical qubits, checks of weight 9 and, by Eq.~\eqref{eq:k}, $k \ge |G|$. Bhardwaj et al.\ also require $\lambda(a_2)$ and $\rho(b_2)$ to be invertible, which gives $k = |G|$. Over $Z_6 \times S_3$ only $123$ of the $595$ choices of $a_2$, and of $b_2$, meet this requirement. We drop it, so our family is much larger and contains codes with $k > |G|$. In this family $d_X$ and $d_Z$ can differ, so the best codes depend on the bias between bit and phase flips.

Once each sum is multiplied by a group element so that it contains the identity, we require the elements of $a_1$ and $a_2$ to generate $G$, that is, every element of $G$ is a product of them and their inverses, and likewise the elements of $b_1$ and $b_2$. This discards every code that splits into disconnected copies of a smaller code~\cite{lin2023}.

\subsection{GFlowNets}
\label{sec:primer}
A GFlowNet~\cite{bengio2021,bengio2023} builds an object one piece per step. A neural network with parameters $\theta$, called the policy, chooses each piece and is trained to produce each complete object with probability proportional to a positive reward:
\begin{equation}
  p_\theta(x) = \frac{R(x)}{\sum_{x'} R(x')},
  \label{eq:gfn-target}
\end{equation}
where $x'$ runs over all complete objects. When there are too many objects to list, this total reward cannot be computed, so it is learned as a scalar $Z_\theta$ alongside the network.

%% file: sections/04-benchmark.tex
\section{Search Space and Oracle}
\label{sec:benchmark}

\subsection{Equivalent Codes and the Canonical Form}
\label{sec:equivalence}
Two CSS codes are equivalent when relabelling the qubits and the checks turns the matrices of one into those of the other, and equivalent codes share $n$, $k$, $d_X$ and $d_Z$. Four operations on a tuple produce an equivalent tuple: reordering the elements within a sum, translating a sum (multiplying all its elements by one group element), renaming the elements of $G$ by an automorphism (a one-to-one renaming that respects the multiplication table), and conjugating the $g_i$ alone or the $h_i$ alone ($g_i \to g g_i g^{-1}$ for one $g$). The canonical form of a tuple undoes only reordering and translation. Two distinct canonical forms can still be equivalent codes, and a graph-isomorphism program decides equivalence exactly on the Tanner graphs. The mirror image of a code swaps $H_X$ and $H_Z$, so that $d_X$ and $d_Z$ trade places, and we count a code and its mirror image as two codes.

\subsection{Oracle and Reference Sample}
\label{sec:oracle}
\label{sec:keys}
The oracle returns the score $s(x)$ of the code of a tuple $x$, with higher scores for better codes.
Every search method gets its scores only from the oracle, which first replaces the tuple by its canonical form. A form already scored in the run is answered from memory at no cost, whereas a new form is a paid call, so the cost of a method is the number of paid calls it makes.

The score assumes that each qubit suffers a bit flip with probability $p_X$ and, independently, a phase flip with probability $p_Z$.
The noise bias $\gamma = \log_{10}(p_X/p_Z)$ says which flip is more likely, while the rarer flip is fixed at $10^{-3}$.
A code of phase-flip distance $d_Z$ corrects every pattern of fewer than $d_Z/2$ phase flips, so the lightest pattern that defeats it has probability about $p_Z^{d_Z/2}$, and the same holds for bit flips with $d_X$ and $p_X$.
The score, which we call the failure exponent, counts the orders of magnitude by which the more likely of these two failures is suppressed:
\begin{equation}
\begin{aligned}
  s(x, \gamma) &= \min(c_X d_X,\; c_Z d_Z), \\
  c_X &= \tfrac{1}{2}\log_{10}\frac{1}{p_X}, \quad c_Z = \tfrac{1}{2}\log_{10}\frac{1}{p_Z}.
\end{aligned}
\label{eq:exponent}
\end{equation}
A score of $10$, for example, means that the lightest uncorrectable error has probability about $10^{-10}$.
We report results at $\gamma = \pm 1$, where the likelier flip has probability $10^{-2}$ and coefficient $1.0$, and the rarer one has coefficient $1.5$. We write $s(x)$ when the bias is fixed.

The threshold for a top-scoring code is the highest score in the uniform reference sample, the $666$ of $1{,}000$ uniform draws whose elements generate $G$, each with exact distances: $10.0$ at $\gamma = \pm 1$, reached by $6$ of the $666$. A top-scoring code thus has $d_X \ge 10$ and $d_Z \ge 7$ at $\gamma = +1$, and the reverse at $\gamma = -1$. At $\gamma = 0$, by contrast, $116$ of the $666$ share the sample's highest score, so this threshold would not separate methods. The best codes certified by search runs have $d_X = d_Z = 10$, so they are $[[180, k, 10]]$ codes with $k \ge 36$.

Exact distances come from what we call the SAT ladder, an established use of a satisfiability solver~\cite{chen2026sat}.
For $w = 1, 2, \ldots$ in turn, the ladder asks whether a logical operator of weight at most $w$ exists, and a question left undecided within its budget stops the ladder at a lower bound.
Two such ladders, the exact ladders, give $d_X$ and $d_Z$.
They are expensive, so every new code first gets a cheap and slightly optimistic bound score. It uses $d_A$ and $d_B$, the distances of the classical codes with check matrices $(\lambda(a_1) \mid \lambda(a_2))$ and $(\rho(b_1) \mid \rho(b_2))$, computed by ladders of the same kind, the classical ladders. A lightest codeword of the first, placed on two of the five blocks of qubits, is a phase-flip pattern that no $X$ check detects, so $d_Z \le d_A$ unless it is a stabilizer, and likewise $d_X \le d_B$. The bound score is Eq.~\eqref{eq:exponent} with $d_B$ and $d_A$ in place of $d_X$ and $d_Z$.
At $\gamma = \pm 1$, the same paid call also runs the exact ladders whenever the bound score reaches the threshold or a classical ladder is left undecided. A code on which they have run is certified, and it counts as top-scoring when the score they prove reaches the threshold. A ladder that stops early, as happens for $0.7$ per cent of certified codes, or a classical bound that fails can leave a good code uncounted, but neither can make a bad code count.
On one CPU core a bound score takes $0.8$~s in the median and $1.5$~s in the mean, whereas the exact ladders take about $10$~s on a uniform draw and about $2$~min on a certified code.

%% file: sections/05-policy.tex
\section{GFlowQEC}
\label{sec:policy}
\begin{figure*}[t]
  \centering
  \includegraphics[width=\textwidth]{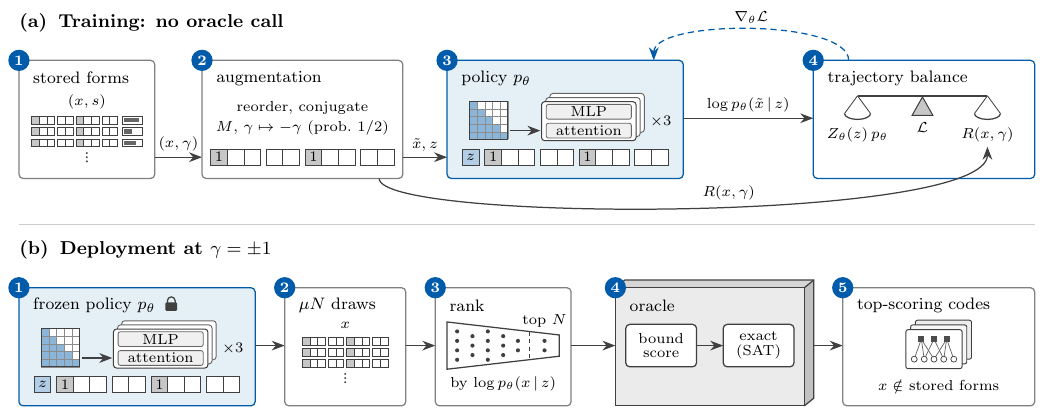}
  \caption{GFlowQEC ($n = 180$). Rows of slots are tuples, with grey slots fixed to the identity, and Tanner graphs are codes. (a) Training, with no oracle call: (1) a stored form $x$ with its score $s$ and a bias $\gamma$, which give its reward $R(x, \gamma)$, (2) augmented and, with probability $1/2$, mirrored by $M$ with $\gamma \mapsto -\gamma$, (3) read by the transformer (three blocks, causal mask shaded), which takes the bias as a condition $z$ on its start token, (4) the trajectory-balance loss, weighing $R(x, \gamma)$ against $Z_\theta(z)\, p_\theta$, and its gradient step (dashed). (b) Deployment: (1) the frozen policy at $\gamma = \pm 1$, (2) $\mu N$ drawn forms (Sec.~\ref{sec:deploy}), (3) the $N$ with the highest $\log p_\theta(x \mid z)$, (4) the oracle's bound score and SAT certification, (5) the top-scoring codes outside the stored forms.}
  \label{fig:pipeline-offline}
\end{figure*}

GFlowQEC's policy is a causal transformer~\cite{vaswani2017} that builds the tuple $x$ one position, or slot, at a time, and its training target is
\begin{equation}
  p_\theta(x) \propto R(x) = \max(s(x), 0.01)^{\beta},
  \label{eq:proportional}
\end{equation}
where $p_\theta(x)$ is the probability that the policy builds $x$, the score $s(x)$ is clamped at $0.01$, and the exponent $\beta > 0$ sets how strongly the policy favours high scores. Fig.~\ref{fig:pipeline-offline} shows its training and deployment.

\subsection{Slot-by-Slot Construction}
\label{sec:emission}
The tuple has $T = 10$ slots, which the policy fills in the order $g_1, \ldots, g_5$, then $h_1, \ldots, h_5$.
We write $x_t$ for the element in slot $t$, $x_{<t}$ for the elements before it, and $P_F(x_t \mid x_{<t}; \theta)$ for the policy's probability of $x_t$.
Slots $g_1$ and $h_1$ are fixed to the identity. By translation this loses no code up to equivalence, and it puts the identity in every sum. To keep the elements of each sum distinct, a mask restricts every other slot $t$ to its legal set, which is the group without the identity and without the elements already placed in the same sum,
\begin{equation}
  \mathcal{A}_t(x_{<t}) = G \setminus \bigl( \{1\} \cup \{ x_{t'} : t' < t,\ t' \text{ in the same sum as } t \} \bigr).
  \label{eq:mask}
\end{equation}
A tuple is legal when $g_1 = h_1 = 1$ and every other element lies in its legal set, so there are $((|G|-1)(|G|-2))^4$ legal tuples.
A cheap test, the generation check, decides without an oracle call whether the elements generate $G$, and a tuple that fails it is never sent to the oracle.

\subsection{Objective and the Choice of \texorpdfstring{$\beta$}{beta}}
\label{sec:objective}
Because a partly filled tuple can be reached only by filling its slots in order, the graph of partial tuples, whose edges each add one element, is a tree.
The probability of producing $x$ is then the product of the slot probabilities,
\begin{equation}
  p_\theta(x) = \prod_{t=1}^{T} P_F(x_t \mid x_{<t}; \theta).
  \label{eq:tree}
\end{equation}
Trajectory balance~\cite{malkin2022} then requires $Z_\theta\, p_\theta(x) = R(x)$, and we minimise the squared gap between the logarithms of its two sides,
\begin{equation}
  \mathcal{L}(x) = \Bigl( \log Z_\theta + \sum_{t=1}^{T} \log P_F(x_t \mid x_{<t}; \theta) - \log R(x) \Bigr)^2,
  \label{eq:loss-tree}
\end{equation}
with $\log$ the natural logarithm.

We set the exponent to $\beta = 8$ before training by a fixed rule: it is the smallest of $1, 2, 4, 8, 16, 27, 40$ and $64$ at which the best tenth of the uniform reference sample and its mirror images, ranked by clamped score, holds at least four fifths of their total reward.

\subsection{Training Data}
\label{sec:cache}
\label{sec:symmetry}
The policy trains only on scores that $42$ earlier runs of several search methods had already paid for, so training itself makes no oracle call. Those runs made $126{,}000$ paid calls covering $116{,}972$ distinct canonical forms, the stored forms, and include the tuned PT runs behind seeds 0 to 2 of Table~\ref{tab:sum180}.
Each form is stored with the two distances the oracle returned, which give its score at any bias (Eq.~\eqref{eq:exponent}): exact distances for the $6{,}770$ certified codes, apart from $45$ whose ladder stopped early at a lower bound, and the bounds $d_A$ and $d_B$ for the other $110{,}202$.

During training, each drawn tuple is replaced by a random equivalent tuple $\tilde{x}$ with the same reward, obtained by shuffling the two free elements of each sum and conjugating the $g_i$ by one random group element and the $h_i$ by another, never by translation.
With probability one half, a training tuple is also replaced by its mirror $M(x) = (h_1^{-1}, \ldots, h_5^{-1}, g_1^{-1}, \ldots, g_5^{-1})$, which builds a code equivalent to the mirror image of $x$, so it scores at $-\gamma$ what $x$ scores at $\gamma$ and trains at the opposite bias with the same reward.
A single network serves both biases by taking the bias as an input, a condition vector $z$, and each bias has its own learned normaliser $Z_\theta(z)$.
We train one model per random seed, ten in all, at $\gamma = \pm 1$, the biases at which they are deployed, and then freeze them.
The transformer has three blocks of width $128$ with four heads ($605{,}863$ parameters) and trains with Adam~\cite{kingma2015} for $315{,}280$ updates of $256$ tuples, about $690$ passes over the stored forms.

\subsection{Deployment}
\label{sec:offline}
\label{sec:offline-training}
\label{sec:deploy}
The frozen policy proposes codes at $\gamma = +1$ or $\gamma = -1$. It draws until it holds $\mu N$ distinct canonical forms, where $N = 1{,}000$ is the budget of paid calls and $\mu = 10$ or $100$ is an oversample factor, and discards without an oracle call every draw that fails the generation check or repeats a form. It then ranks the forms by $\log p_\theta(x \mid z)$, the log-probability of the tuple as drawn, which stands in for the score: at zero loss $\log p_\theta(x \mid z) = \beta \log \max(s(x), 0.01) - \log Z_\theta(z)$, so at a fixed bias it orders the stored tuples by their stored scores. It pays one oracle call for each of the $N$ highest. Draws are not checked against the stored forms, so a form that is sent may already be stored, and it is still paid for.
We chose this ranking after reading earlier runs, in which paying for the draws in the order drawn had trailed an earlier, smaller supervised screen per paid call, in every seed at both biases. Ranking found more than five and a half times as many top-scoring codes as paying in draw order.

%% file: sections/08-evaluation-protocol.tex
\section{Experimental Setup}
\label{sec:protocol}
\label{sec:grading-units}

\subsection{Comparison Methods}
\label{sec:methods}
Every method uses the same oracle, search space and budget, and runs are paired by random seed.
\begin{itemize}
\setlength{\itemsep}{0pt}
\item \emph{Tuned parallel tempering} (tuned PT)~\cite{swendsen1986,geyer1991,hukushima1996} runs eight chains of local search at different temperatures, each move redrawing one slot uniformly, with Metropolis acceptance~\cite{metropolis1953} and exchanges between the chains. It uses the better of two temperature ranges on its own runs, $0.25$ to $2.0$, a choice that favours it. A variant starts its chains at stored forms rather than at uniform draws, and those starting forms never count toward its results.
\item The \emph{supervised screen} uses the policy's network as a regressor that predicts a code's score. Unlike the policy, it sees all ten slots at once, and it also reads features that do not change when a sum is translated. It is trained by squared error to predict the scores of the same stored forms, with the policy's augmentation, batches and seeds. Its variant and training length were tuned on a held-out tenth of the stored forms, a choice the policy did not get, after which it was trained on all of them. When deployed, it ranks ten or a hundred times $N$ uniform random candidates and pays for its top $N$.
\item \emph{Soft Q-learning}~\cite{haarnoja2017} trained offline uses the same stored scores, network, training steps and deployment rule as GFlowQEC. It shares the GFlowNet's optimal policy (Sec.~\ref{sec:related}), so it tests whether the training objective matters.
\item \emph{MAP-Elites}~\cite{mouret2015}, an evolutionary quality-diversity algorithm, keeps the best code found in each bin of a grid of cheap code properties, and we tuned it over two grids, five bins of $d_A - d_B$ alone or crossed with four bins of $k$.
\item \emph{Bayesian optimisation} (BO) fits a Mat\'ern $5/2$ Gaussian process to a one-hot encoding of the tuple and, after $200$ uniform draws, pays for the best of $2{,}000$ uniform candidates by expected improvement. It is not the method of Chengyu et al.~\cite{chengyu2026}, who use a learned graph embedding and a different decoder, noise rate and code family.
\item \emph{Uniform sampling} pays for uniform random tuples that pass the generation check.
\end{itemize}
We also ran PPO~\cite{schulman2017} at two entropy weights, MAP-Elites on the twenty-bin grid, and soft Q-learning and Munchausen DQN~\cite{vieillard2020} trained during the search, and each found fewer top-scoring codes than tuned PT in every seed at both biases.

\subsection{Measures}
\label{sec:lenses}
The main measure is the number of certified top-scoring codes in a run's first $1{,}000$ paid calls.
A paid code is novel when its canonical form is not among the stored forms. For a method trained on or started from the stored forms, only novel codes count, whereas the other methods never saw the stored forms, so all their codes count.
Because a novel code can still be equivalent to a stored code or its mirror image, we also give the inequivalent count, which drops such codes and, for every method, counts each equivalence class once.

To compare two methods we use their pooled ratio, which divides their counts, each summed over the seeds both methods ran, with the 95 per cent log-ratio interval of Katz et al.~\cite{katz1978}.
For counts $\kappa_1$ and $\kappa_2$ out of $\nu_1$ and $\nu_2$ paid calls, the interval is the ratio times $e^{\pm 1.96\,\omega}$, with $\omega^2 = 1/\kappa_1 - 1/\nu_1 + 1/\kappa_2 - 1/\nu_2$.
We say that the two are not separated when the interval contains $1$.
It treats every paid call as independent, which leaves out the variation between seeds and within a search chain, so we also give the number of seeds in which one method leads.

%% file: sections/09-results.tex
\section{Results}
\label{sec:results}

\label{sec:res-offline}
\begin{table*}[!t]
\centering
\caption{Results at $180$ physical qubits: certified top-scoring codes in each run's first $1{,}000$ paid calls, mean over ten seeds rounded to whole codes, at noise bias $\gamma = +1$ and $\gamma = -1$. In the first group of rows only novel codes count. Inequivalent: Sec.~\ref{sec:lenses}. Ratios in the text pool the seeds both methods ran.}
\label{tab:sum180}
\footnotesize
\setlength{\tabcolsep}{3pt}
\begin{tabular}{@{}>{\raggedright\arraybackslash}p{\resmethw}*{4}{>{\centering\arraybackslash}p{\resdatw}}@{}}
\toprule
 & \multicolumn{2}{>{\centering\arraybackslash}p{\resheadw}}{Certified top-scoring codes} & \multicolumn{2}{>{\centering\arraybackslash}p{\resheadw}@{}}{Inequivalent} \\
\cmidrule(lr){2-3}\cmidrule(l){4-5}
Method & $\gamma = +1$ & $\gamma = -1$ & $\gamma = +1$ & $\gamma = -1$ \\
\midrule
\multicolumn{5}{@{}l}{\textit{Trained on, or started from, the stored forms}} \\
\midrule
GFlowQEC & $313$ & $317$ & $221$ & $222$ \\
\rowrule
Supervised screen & $87$ & $84$ & $62$ & $59$ \\
\rowrule
Soft Q-learning, trained offline & $317$ & $364$ & $225$ & $259$ \\
\rowrule
Tuned PT from the stored forms & $119$ & $123$ & $79$ & $84$ \\
\midrule
\multicolumn{5}{@{}l}{\textit{Searching without stored data}} \\
\midrule
Tuned PT & $97$ & $84$ & $92$ & $79$ \\
\rowrule
MAP-Elites, five bins & $103$ & $92$ & $88$ & $74$ \\
\rowrule
Uniform sampling & $12$ & $8$ & $12$ & $8$ \\
\rowrule
BO, one-hot encoding & $50$ & $53$ & $49$ & $52$ \\
\bottomrule
\end{tabular}
\end{table*}
Table~\ref{tab:sum180} gives the counts, with GFlowQEC and the supervised screen at $\mu = 10$.
In every paired seed and at both biases, GFlowQEC (below, the policy) found more top-scoring codes than any other method in the table except soft Q-learning trained offline. Each pair of numbers below is at $\gamma = +1$ and $\gamma = -1$, in that order.

\paragraph{Against tuned PT.}
On inequivalent codes, tuned PT is the strongest method without the stored data, although MAP-Elites with five bins found slightly more before equivalent codes were merged.
The policy found $2.4$ $[2.2, 2.6]$ and $2.8$ $[2.6, 3.0]$ times as many inequivalent top-scoring codes as tuned PT, and more than twice as many in $19$ of its $20$ runs.
Starting tuned PT from the stored forms does not close the gap, since on the same count the policy still found $2.8$ and $2.7$ times as many.

\paragraph{Against soft Q-learning.}
Soft Q-learning trained offline found $1.01$ and $1.15$ times the policy's novel count, not separated at $\gamma = +1$ and ahead at $\gamma = -1$, where the interval $[1.10, 1.19]$ excludes $1$. It led in $7$ and $8$ of the ten seeds.

\paragraph{Against the supervised screen.}
At $\mu = 10$ the policy found $3.6$ and $3.8$ times as many top-scoring codes as the screen, and led in each of ten seeds.
When both oversampled at $\mu = 100$, the screen found more in each of ten seeds, and the policy found $0.45$ $[0.44, 0.47]$ and $0.62$ $[0.61, 0.64]$ times its count.
Still, two controls show that the policy's strength lies in its draws.
Ranking the screen's uniform candidates at $\mu = 10$ instead of its own draws, the policy found a fifth as many codes as from its own draws, and fewer than the screen.
Ranking the policy's draws instead of uniform candidates at $\mu = 100$, the screen found $1.27$ and $1.30$ times its own count, more in every seed, so better draws and better ranking combine.

\paragraph{Origin of the codes.}
At $\gamma = +1$ the cap $d_X \le d_B$ depends only on the $h_i$, so we count two top-scoring codes, novel or not, as one neighbourhood when their $h_i$ agree up to conjugating every element by one group element, and at $\gamma = -1$ the $g_i$ play this role.
In each of the policy's runs, $87$ to $93$ per cent of its neighbourhoods also hold a stored top-scoring code, so the policy mostly finds new partners for good halves that the stored data already held.

\paragraph{Cost.}
The policy's paid calls ran the exact ladders $3.5$ and $4.2$ times as often as tuned PT's, and these ladders dominate the cost of a paid call here.
On seeds 0 to 2, per CPU second of exact ladders, the policy and tuned PT found top-scoring codes at the same rate, within five per cent at both biases, and on the inequivalent counts tuned PT was ahead. On seeds 3 to 9 the two ran on different machines, so we do not compare their CPU seconds.

%% file: sections/10-discussion.tex
\section{Discussion}
\label{sec:discussion}

A trained proposer is most likely to pay off when four conditions hold together: good codes are too rare to find by sampling, no quantity that is cheap to compute from the parity checks ranks them, each call to the oracle is expensive, and the cost of training is shared across many searches. At $180$ physical qubits, good codes are not rare enough, which is likely why a screen shown a hundred random codes per paid call overtakes the policy. But the number of logical qubits $k$, the obvious cheap quantity, does not single out the top-scoring codes: sorting the stored scores by $k$ would raise their top-scoring share by at most $1.4$ times. Of the four conditions, only the cost of a call is known to grow, both with the size of the code and under circuit-level noise. And searches already run at $600$ to $1{,}500$ physical qubits~\cite{liu2026}, past the few hundred physical qubits where exhaustive catalogues stop~\cite{liang2025}.

\paragraph{Limitations.}
\begin{itemize}
\setlength{\itemsep}{0pt}
\item The gain is in how many good codes a budget buys, not in the best code found. We show no code better than those found by Bayesian optimisation, reinforcement learning or LLM-driven searches. Nor does a good code at this size need a trained policy, since uniform sampling found about ten top-scoring codes per $1{,}000$ paid calls (Table~\ref{tab:sum180}).
\item The lead is counted per paid call, with nothing charged for the stored data (the $126{,}000$ paid calls of Sec.~\ref{sec:cache}, against $1{,}000$ per run) or for the policy's training, and it disappears per CPU second of exact ladders (Sec.~\ref{sec:res-offline}).
\item The score is computed from distances alone (Eq.~\eqref{eq:exponent}), with no decoder. We have not measured whether circuit-level noise, with faulty check measurements and a real decoder, reorders the codes.
\item We use the policy as a proposer that ranks its own draws and have not tested whether it samples in proportion to its reward.
\end{itemize}

%% file: sections/11-future-work.tex
\section{Conclusion}
\label{sec:future}
Trained once on stored scores and used frozen at $180$ physical qubits, GFlowQEC found more than twice as many top-scoring codes per paid call as tuned PT, counting only codes inequivalent to its training data. The gain lies in the candidates, not the training loss, since soft Q-learning did as well or better and a screen did best when it ranked the policy's draws.

\paragraph{Future work.}
The next step toward choosing a code for a device is to re-score the top-scoring codes at $180$ physical qubits under circuit-level noise and see whether their ranking survives. The proposer would then be graded at that cost, against the method of Chengyu et al.~\cite{chengyu2026} at the budget that method is built for, a few hundred calls to an expensive oracle. A policy conditioned on the measured noise through multi-objective GFlowNets~\cite{jain2023} could propose a new set of good codes after each calibration, without running a new search.

The setting we aim for is a quantum compiler that chooses the code itself. A strategic plan for neutral-atom quantum computing~\cite{menssen2026} calls for the compiler to select codes from the application and the hardware specification, and names generative neural networks among the methods that could find codes suited to a device's noise and operational constraints. GFlowQEC fits this role as a conditional generator, in which one policy would answer every request. The hard constraints of a request, such as the device's connectivity, the check weight or a floor on $k$, would be built into the mask of Eq.~\eqref{eq:mask}, and its noise model would enter through the condition $z$.

A single policy could also search across code families if each action chose one level of a construction in turn: the family, the group, the base matrices and the sums that fill them. Liu and Marquardt~\cite{liu2026} let a language model mutate a three-level grammar of this kind, whereas ours would be one trained policy choosing every level.